\documentclass[10pt]{amsart}

\usepackage{color}
\usepackage{amsmath}
\usepackage{amssymb}
\usepackage{float}
\usepackage{graphicx}

\newtheorem{proposition}{Proposition}[section]

\newsavebox{\savepar}

\def\G{\mathcal{G}}

\def\N{\mathcal{N}}

\def\E{\mathcal{E}}

\def\epsilon{\varepsilon}
\def\leq{\leqslant}

\def\c{{\bf c}}

\def\int{{\rm int}}

\def\G{\mathcal{G}}
\def\SG{\mathcal{S}}

\def\N{{\bf N}}

\def\E{{\bf E}}
\def\X{{\bf X}}

\def\BSigma{{\bf \Sigma}}

\def\Phat{\hat{P}}
\def\phat{\hat{p}}

\newcommand{\rem}[1]{}

\begin{document}
\title{Markov Chain Aggregation for Simple Agent-Based Models on Symmetric Networks: The Voter Model\\\vspace{12pt}
Extended Version of:\\ Markov Projections of the Voter Model (1209.3902v1) }
\author{Sven Banisch$^{a,c,d}$ and Ricardo Lima$^{b,d}$}

\maketitle

\markboth{Banisch/Lima}{Markov Chain Aggregation for Simple Agent-Based Models on Symmetric Networks: The Voter Model}

\begin{center}
\begin{minipage}{13truecm}
\begin{small}

($^{A}$) Max Planck Institute for Mathematics in the Sciences (Germany)
\footnote{Corresponding author:
Sven Banisch,
MPI-MIS,
Inselstrasse 22, D-04103 Leipzig, Germany\\
eMail: sven.banisch@UniVerseCity.de
 }\\
($^{B}$) Dream \& Science Factory, Marseilles (France)\\
($^{C}$) Mathematical Physics, Bielefeld University (Germany)\\
($^{D}$) Institute for Complexity Sciences (ICC), Lisbon (Portugal)\\

\end{small}
\end{minipage}
\end{center}

\vskip 2truecm



\begin{abstract}
For Agent Based Models, in particular the Voter Model (VM),  a general framework of aggregation is developed which exploits the symmetries of the agent network $G$.
Depending on the symmetry group $Aut_{\omega} (N)$ of the weighted agent network, certain ensembles of agent configurations can be interchanged without affecting the dynamical properties of the VM.
These configurations can be aggregated into the same macro state and the dynamical process projected onto these states is, contrary to the general case, still a Markov chain.
The method facilitates the analysis of the relation between microscopic processes and a their aggregation to a macroscopic level of description and  informs about the complexity of a system introduced by heterogeneous interaction relations.
In some cases the macro chain is solvable.
\end{abstract}


\section{Introduction} 
\label{introduction}

Agent--based models (ABMs) are from a formal point of view Markov chains \cite{Izquierdo2009,Banisch2012son}.
However, there is not too much to learn from this fact alone as the state space of the chains corresponding to those models must be conceived of as the set of all possible configurations of the entire agent system.
The resulting Markov chain is too big for a tractable analysis of a model and therefore a considerable reduction of the state space is required.

In practice, the models are often re--formulated in terms of an aggregate variable (such as magnetization) so that all micro configurations with the same aggregate value are agglomerated in a single entity.
However, the dynamical process at a coarse--grained level obtained in this way is, in general, no longer a Markov chain, because mere aggregation over agent attributes is insensitive to microscopic heterogeneity and some information about the dynamical details are lost.
In fact, as shown in \cite{Banisch2012son} in the context of opinion dynamics, Markovianity is preserved by the macro description only in the case of uniform interaction probabilities between the agents, that is, for homogeneous mixing.
In other cases, therefore, the results are usually considered as an approximation and compared to simulations.
Another possibility, however, is to refine the aggregation procedure such that the Markov property is not lost and try to solve the refined problem.
In this paper a systematic approach to aggregation is developed which exploits all the dynamical redundancies that have its source in the agent network on which the model is implemented.

Though it has often been recognized that ABMs may be conceived as stochastic dynamical systems or Markov chains \cite[]{Epstein1996,Laubenbacher2009,Izquierdo2009,Page2012}, the aggregation techniques developed for these systems have not yet been applied to ABMs.

On the other hand, the mathematical analysis of the relation between microscopic processes and a their aggregation to a macroscopic level of description is at the heart of statistical mechanics and has recently received some attention in the complex systems literature.
In \cite{Shalizi2003}, the particular role of Markovianity in the definition or identification of macroscopic observables is emphasized.
Based on that, \cite{Goernerup2008} develops a method to derive a Markovian state space aggregation on the basis of an information--theoretic view on time series data (typically created by some simple model).
These and related concepts for the level identification in complex dynamical systems are compared and related in \cite{Pfante2013} that emphasize the particular role played by commutativity of aggregation and dynamics.

Moreover, there is a considerable amount of work done in Markov chain theory dedicated to lumpability.
Consider a Markov chain on the state space $\BSigma$ along with a partition $\X$ of that state space.
Lumpability is about whether or not the projection of original process onto $\X$ preserves Markovianity.
Notice that such a situation naturally arises if the process is observed not at the micro level of $\BSigma$, but rather in terms of a measure $\phi$ on $\BSigma$ by which all states in $\BSigma$ that give rise to the same measurement are mapped into the same aggregate set $X_k \in \X$ (also referred to as macro states).
Notice also that for a given chain there may be several lumpable partitions.
In this paper we follow \cite{Kemeny1976} and refer to \cite{Burke1958,Rosenblatt1959,Rogers1981} for other seminal works on Markovian aggregation.

Most methods for the identification of lumpable partitions rely on the transition matrix of the original chain, among them spectral methods and eigenvector conditions proposed in \cite[]{Barr1977,Meila2001,Takacs2006,Jacobi2008,Goernerup2010}.
For ABMs, however, the state space of the chain becomes huge even for a relatively small number of agents.
Even in the simplest case of binary agent attributes, the size of the Markov chain scales as $2^N$ with the number of agents $N$.
Therefore, another approach is developed which allows us to identify lumpable partitions as a function of the network symmetries (size $N$) instead of the properties of the Markov chain associated to the ABM (size $2^N$).
Deriving aggregate descriptions by starting from of the symmetries of the agent network, our approach is related to the hierarchical approach due to \cite{Buchholz1995} and the idea of symmetric composition in \cite{Hermanns1999}.

\section{The Voter Model Revisited} 
\label{votermodel}

We use the Voter Model (VM) to exemplify our approach (see Refs. \cite{Kimura1964,Clifford1973,Frachebourg1996,Slanina2003,Sood2005,Vazquez2008}).
The VM has its origin in the population genetics literature \cite{Kimura1964}, but due to its interpretation as a model for opinion dynamics it has become a benchmark model in social dynamics.
The reader is referred to Ref. \cite{Korolev2010} for a recent review of related models in population genetics and to Ref. \cite{Castellano2009} for an overview over the social dynamics literature.

Consider a set $\N$ of $N$ agents, each one characterized by an individual attribute $x_i$ which takes a value among two possible alternatives: $x_i \in \{\square,\blacksquare\}, i = 1,\ldots,N$.
Let us refer to the set of all possible combinations of attributes of the agents as configuration space and denote it as $\BSigma = \{\square,\blacksquare\}^N$.
Therefore, if $x \in \bf{\Sigma}$ we write $x = (x_1,\dots, x_i, \dots, x_N)$ with $x_i \in \{\square,\blacksquare\}$.
The agents are placed on a network $G = (\N,\E)$ where $\N$ corresponds to the set of $N$ agents and $\E$ is the set of agent connections $(i,j)$.
In the dynamic process implemented in the VM an agent $i$ is chosen at random along with one of its neighboring agents $j$ according to a probability distribution $\omega(i,j)$. 
If the states ($x_i,x_j$) are not equal already, agent $i$ adopts the state of $j$ (by setting $x_i=x_j$).
Notice that sometimes there is a probabilistic choice whether agent $i$ or $j$ imitates the other, but this does not affect the dynamical behavior in the undirected case with $\omega(i,j) = \omega(j,i)$.
Notice also that there are two possible modes of agent choice referred to as node or link update dynamics and that dealing with a distribution $\omega$ encompasses the two (see Sec. \ref{sec:2Com}).

It is well--known that the VM has two absorbing states corresponding to the configurations with all agents holding the same opinion (consensus).
Analytical results for the mean convergence times have been obtained for the complete graph \cite{Slanina2003,Banisch2012son}, for $d$--dimensional lattices \cite{Cox1989,Frachebourg1996,Liggett1999,Krapivsky2003} as well as for networks with uncorrelated degree distributions~\cite{Sood2005,Vazquez2008}.

\subsection{Micro Dynamics}

\begin{figure}[hbtp]
\centering
\includegraphics[width=.8\linewidth]{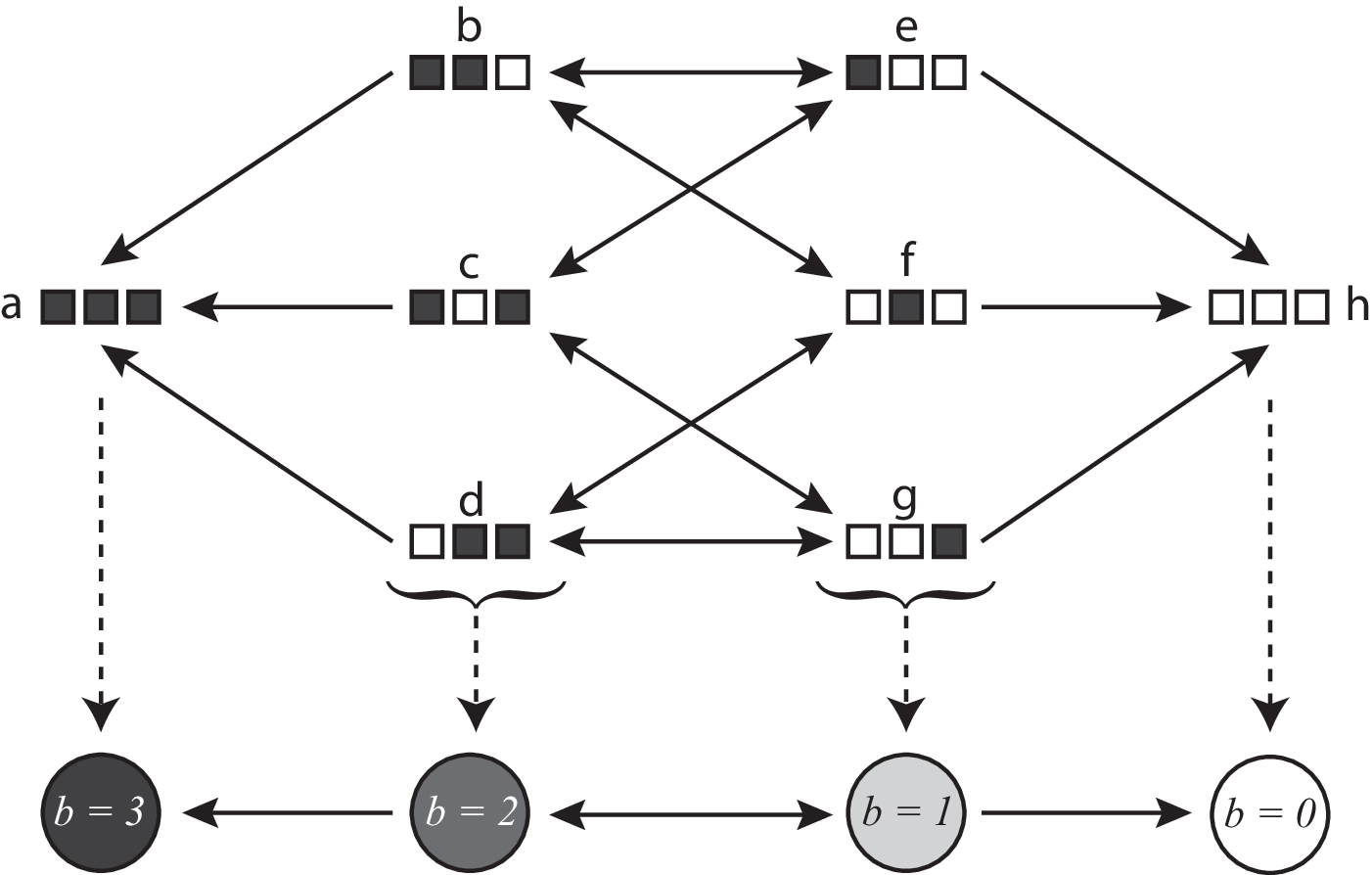}
\caption{The micro chain for the VM with 3 agents and its projection obtained by agglomeration of states with the same number of black agents~$b$. Only exceptionally the projected system will be a Markov chain.}
\label{fig:3Hypercube}
\end{figure}

At the microscopic level of all possible configurations of agents the VM can be seen as an absorbing random walk on the $N$--dimensional hypercube with the two absorbing states $(\blacksquare,\ldots,\blacksquare)$ and $(\square,\ldots,\square)$.
This is shown for the example of three agents in Fig. \ref{fig:3Hypercube}.
Due to the sequential update process, from one interaction event to the other, only one agent changes or the configuration remains as it is (loops are not shown in Fig. \ref{fig:3Hypercube}).
That is, non--zero transition probabilities exist only between configurations that are adjacent in the hypercube graph.
Let us therefore call two agent configurations $y,x \in \BSigma$ adjacent if they differ only with respect to a single agent $i$.
We denote this by $x \stackrel{i}{\sim}  y$.

The VM belongs to a more general class of models in which the update of an agent (say $i$) depends only on its local neighborhood (in this case one other randomly chosen agent $j$).
Therefore (see \cite{Banisch2012son}), the transition probability for a pair of adjacent configurations $x \stackrel{i}{\sim}  y$ can be written as
\begin{equation} 
\Phat(x,y) = \sum\limits_{j: \left(\substack{x_j \neq x_i } \right) }^{} \omega(i,j). 
\label{eq:PhatVM}
\end{equation}
That is, the probabilities associated with the arrows in Fig. \ref{fig:3Hypercube} depend only on the probability $\omega$ with which an agent pair $(i,j)$ is chosen such that their opinions $(x_i,x_j)$ result in the respective transition ($x_i \neq x_j$ in the VM).
Different agent interaction networks have therefore a crucial effect on the transition probabilities of the micro chain.

Notice that, at this level, the dynamics of the model is defined in the configuration space  with $2^N$ states, which seeks to describe the dynamics of each agent in full detail.
Following \cite{Banisch2012son} we refer to this as \emph{micro dynamics}.
As the number of states increases as $2^N$ with the number of agents, it is not convenient to study the model dynamics at the micro level.
It is often more convenient to reformulate the dynamics in terms of a macroscopic aggregate variable.
In the VM, the number $b$ of black agents $i$ with $x_i = \blacksquare$ or the respective density are natural choices.
Fig. \ref{fig:3Hypercube} illustrates how several micro configurations with the same $b$ are taken together to form a single macro state.
Notice that in the hypercube terminology $b$ corresponds to the Hamming weight of a configuration.

\section{Markovian Aggregation}
\label{markovaggregation}

The main question addressed by this paper concerns the conditions under which an aggregation of micro states leads to a Markov chain at the resulting macro level.
More precisely, we make use of a theorem in Markov chain theory that gives necessary and sufficient conditions for partitioning the state space of Markov chains in such a way that the micro process projected onto that partition is still a Markov chain, \cite{Kemeny1976} Thm. 6.3.2.
Let $\phat_{x Y} = \sum\limits_{y \in Y}^{} \hat{P}(x,y)$ denote the conjoint probability for $x \in \BSigma$ to go to the set of elements $y \in Y$ where $Y \subseteq \BSigma$ is a subset of the configuration space.
According to \cite{Kemeny1976}, Thm. 6.3.2, the micro chain $(\BSigma,\Phat)$ is \emph{lumpable} with respect to a partition $\X =  (X_1,\ldots,X_r)$, iff, for every pair of sets $X_i$ and $X_j$, the probability $\phat_{x X_j}$ is the same for every $x \in X_i$, and these common values form the transition matrix for the lumped chain.
In general it may happen that, for a given Markov chain, some projections are Markov and others not.

Based on \cite{Kemeny1976}, Thm. 6.3.2, we have derived in \cite{Banisch2012son}, Prop. 3.1, a sufficient condition for lumpability which exploits the symmetries of the micro chain $\Phat$.
Namely, to any partition $\X$ we can associate a transformation group $\G$ acting on $\BSigma$ the orbits of which generate $\X$.
Then, a sufficient condition for lumpability is that the transition matrix $\Phat$ is symmetric with respect to $\G$, that is, $\Phat(x,y) = \Phat(\hat{\sigma} (x),\hat{\sigma} (y))$ for any $\hat{\sigma} \in \G$.
This means that the identification of a lumpable partition can be based on the symmetries of the micro chain. 
However, it is worth to notice that in this paper lumpability conditions are stated in terms of the symmetries of the agent network which is much simpler than the dynamical graph on the configuration space where the aggregation process (lump) is achieved. 


Let $Aut_{\omega}(N)$ be the subgroup of the permutations $\sigma$ acting on the set $\N$ of agents such that $\omega(\sigma i, \sigma j) = \omega( i, j)$ for all $i,j \in \N$.
To each $\sigma \in Aut_{\omega}(N)$ we associate a $\hat{\sigma}$ which is a bijection on the configuration space $\BSigma$.
If $x\in \BSigma$ with $x= (x_{1}, \dots,  x_{ i}, \dots,  x_{N})$ then
\begin{equation}
\label{hatsigma}
\hat{\sigma} (x) = (x_{\sigma 1}, \dots,  x_{\sigma i}, \dots,  x_{\sigma N}).
\end{equation}

We now define a partition of $\BSigma$ using  $Aut_{\omega} (N)$.
Let us denote as $\G_{\omega} = \{\hat{\sigma} : \sigma \in  Aut_{\omega}(N)\}$ the group of transformations $\hat{\sigma}$ associated to the $\sigma \in Aut_{\omega} (N)$.
The action of $\G_{\omega}$ on $\BSigma$ induces a partition $\mathcal{M}_{\omega}$ of the original state space. 
Accordingly, two configurations $x, x^{\prime} \in \BSigma$ are in the same atom of the partition $\mathcal{M}_{\omega}$ $iff$ there is a $\sigma \in Aut_{\omega} (N)$ such that $ x^{\prime} =  \hat{\sigma} (x)$.
Clearly this is an equivalence relation and we shall refer to two configurations $x$ and $x'$ as \emph{macroscopically equivalent} if they belong to the same atom $X$ of the partition $\mathcal{M}_{\omega}$.


We then have the following:
\begin{proposition}
\label{propositionlumpability}
The partition $\mathcal{M}_{\omega}$ is lumpable for the agent model on $\BSigma$ with agent choice based on $\omega$ and therefore the corresponding projected process is a Markov chain.
\end{proposition}

\begin{proof}
Consider the distribution of interaction probabilities $\omega$, its permutation group of symmetries $Aut_{\omega}(N) = \{\sigma : \omega(\sigma i, \sigma j) = \omega( i, j), \forall i,j \in \N\}$ and the associated transformation group $\G_{\omega} = \{\hat{\sigma} : \sigma \in  Aut_{\omega}(N)\}$.
Suppose we know (at least) one configuration (the generator) $x^k \in \BSigma$ for each $X_k \subset \BSigma$ and construct the partition $\mathcal{M}_{\omega} = (X_1,\ldots,X_k,\ldots)$ by 
\begin{equation}
X_k = \G_{\omega}\circ x^k = \bigcup_{\hat{\sigma} \in \G_{\omega}}^{} \hat{\sigma}(x^k).
\label{eq:construction}
\end{equation}
Following \cite{Banisch2012son},  Prop. 3.1, a sufficient condition for lumpability of $(\BSigma,\Phat)$ with respect to $\mathcal{M}_{\omega}$ is the symmetry of the microscopic transition matrix $\Phat$ with respect to $\G_{\omega}$.
That is, we have to show that 
\begin{equation}
\Phat(x,y) = \Phat(\hat{\sigma} (x),\hat{\sigma} (y)), \forall \hat{\sigma} \in \G_{\omega}.
\label{eq:symmetry}
\end{equation}
For the case that $x \stackrel{i}{\sim}  y$ we know that $x_j =y_j$ for all $j$ except $i$ and that the transition requires the choice of an edge $(i,\ldotp)$.
Denoting $x_i = s$ and $y_i = \bar{s}$ we rewrite Eq. (\ref{eq:PhatVM}) as
\begin{equation} 
\Phat(x,y) = \sum\limits_{j: x_j = \bar{s}}^{} \omega(i,j).
\label{eq:Conject01}
\end{equation}

If $x \stackrel{i}{\sim}  y$ it is easy to show that $\hat{\sigma} (x) \stackrel{\sigma i}{\sim}  \hat{\sigma} (y)$ and 
we know that $s = \hat{\sigma} (x_{\sigma i}) \neq \hat{\sigma} (y_{\sigma i}) = \bar{s}$.
The transition therefore requires the choice of an edge $(\sigma i,\ldotp)$.
We obtain
\begin{equation} 
\Phat(\hat{\sigma} (x),\hat{\sigma} (y)) = 
\sum\limits_{k: \hat{\sigma}(x_k) = \bar{s} }^{} \omega(\sigma i, k).
\label{eq:Conject02}
\end{equation} 
Given an arbitrary configuration $x$, for any $j$ with $x_j = \bar{s}$ we have a corresponding $k = \sigma j$ with $\hat{\sigma}(x_k) = \bar{s}$ because $x_j = \bar{s} \Leftrightarrow \hat{\sigma} (x_{\sigma j}) = \bar{s}$.
That is, the summations in Eq.(\ref{eq:Conject01}) and (\ref{eq:Conject02}) are equal (i.e., the symmetry condition (\ref{eq:symmetry}) is satisfied) for any $\sigma$ for which $\omega(i,j) = \omega(\sigma i,\sigma j)$.
This is true by the definition of $Aut_{\omega}(N)$ for all permutations $\sigma \in Aut_{\omega}(N)$.
Since the transition matrix $\Phat$ is symmetric with respect the associated transformation group $\G_{\omega}$, the partition $\mathcal{M}_{\omega}$ generated by the action of $\G_{\omega}$ on $\BSigma$ is lumpable.
\end{proof}

The reason for which the Markov property is preserved by the projection of the micro chain onto the partition $\mathcal{M}_{\omega}$ is that micro configurations in the same macro state $X_k \in \mathcal{M}_{\omega}$ contribute in precisely the same way to the evolution of the macroscopic process on $\mathcal{M}_{\omega}$.
Namely, all micro states within the same macro set lead to the same assignment of probabilities for a transition to all other macro sets and therefore no additional information is gained about the future evolution by considering the probabilities of being in the specific micro states or, respectively, the histories that led to them.
This is the main idea of Thm. 6.3.2 in \cite{Kemeny1976} and captured by the notion of macroscopic equivalence.

\section{Examples} 
\label{examples}

\subsection{Complete Graph}

We have shown in previous work \cite{Banisch2012son} that the aggregation illustrated in Fig. \ref{fig:3Hypercube} is lumpable only if the interaction probabilities are uniform.
This corresponds to the VM implemented on the complete graph in which  $\omega(i,j) = 1/N(N-1)$ (or $1/N^2$ if self--choice is allowed).
It is, of course, well--known  that the macro model obtained in terms of $b$ fully describes the evolution of the micro model on the complete graph, but not on other topologies (cf. e.g., \cite{Slanina2003}:3 or \cite{Castellano2009}:601).
Nevertheless, our method sheds light on the (probabilistic) reason for this.
Namely, the complete graph and respectively homogeneous mixing is the only topology for which the automorphism group is the group $\SG_N$ of all permutations of $N$ agents.
In this case, for any two configurations $x,x'$ with equal $b$ there is a $\sigma \in \SG_N$ such that $x = \sigma(x')$.
Hence, an equivalent aggregate value $b$ implies macroscopic equivalence.
The fact that this is only true for complete graph and homogeneous mixing, underlines how restrictive these conditions are.

The associated macro process on the partition $\X = (X_0,\ldots, X_b,\ldots X_N)$ is a simple birth--death random walk with $P(X_{b\pm 1}|X_b) = \frac{b (N-b)}{N^2}$,  $P(X_b|X_{b}) = \frac{b^2 + (N-b)^2}{N^2}$ also known as Moran process \cite{Moran1958}.
In \cite{Banisch2012son} we have derived a closed--form expression for the fundamental matrix of that Markov chain for arbitrary $N$.
Encoding the recurrence and hitting times of the system, this provides all the information to characterize the distribution of convergence times.
For instance, for the VM starting in a state with $b$ black agents the mean time to reach consensus is given by
\begin{equation}\label{eq:tauk}
	\tau_b = N \left[  \sum^{b-1}_{j=1} \frac{(N-b)}{(N-j)}  + 1 + \sum^{N-1}_{j= b+1} \frac{b }{j}\right],
\end{equation}
and the respective variance by
\footnotesize
\begin{eqnarray}
\label{eq:tupsilonk}
\upsilon_k =  2N^2 (N-k)\left[ \sum^{k-1}_{i=1} \frac{1}{(N-i)} \left( \sum^{i-1}_{j=1} \frac{(N-i)}{(N-j)} + 1 +  \sum^{N-1}_{j=i+1} \frac{i} {j} \right) \right] 
+ (2N-1) N \left( \sum^{k-1}_{j=1} \frac{(N-k)} {(N-j)} + 1 + \sum^{N-1}_{j=k+1} \frac{k}{j} \right) +\\\nonumber
 + 2 N^2 k \left[ \sum^{N-1}_{i=1} \frac{1}{k+i} \left( \sum^{k+i-1}_{j=1} \frac{(N-k-i)} {(N-j)} + 1 +  \sum^{N-1}_{j=k+i+1} \frac{k+i} {j} \right) \right] 
- N^2 \left( \sum^{k-1}_{j=1} \frac{(N-k)}{N-j} + 1 + \sum^{N-1}_{j= k+1} \frac{k}{j}\right)^2
\end{eqnarray}
\normalsize
We also considered in \cite{Banisch2012son} the VM with more opinions and confirm by lumpability arguments that the convergence times are as in the binary case, which was shown previously in Ref. \cite{Slanina2003}.

\begin{figure}[htbp]
	\centering
		\includegraphics[width=1.0\linewidth]{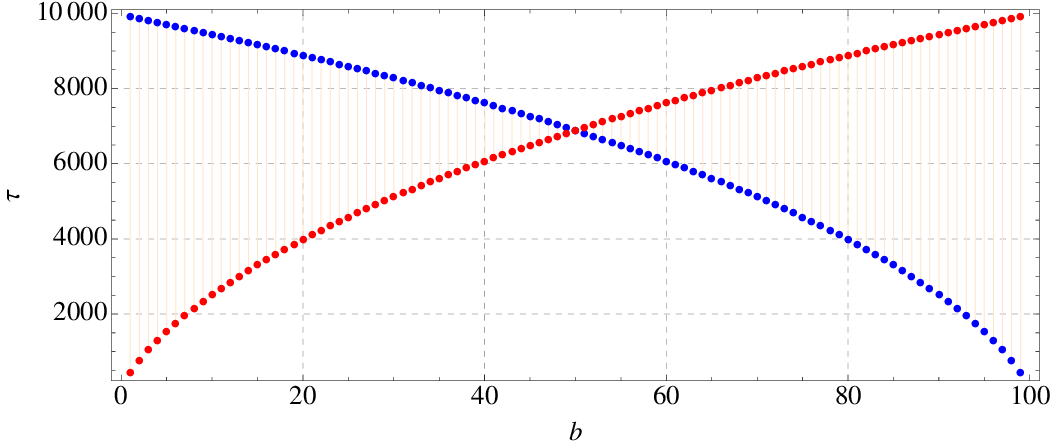}
	\caption{Mean convergence times to $X_{0}$ (red) and $X_{N}$ (blue) for $N = 100$ as a function of the initial number of black agents $b$.}
	\label{fig:TauIndependent}
\end{figure}

\begin{figure}[h]
	\centering
\includegraphics[width=0.6\linewidth]{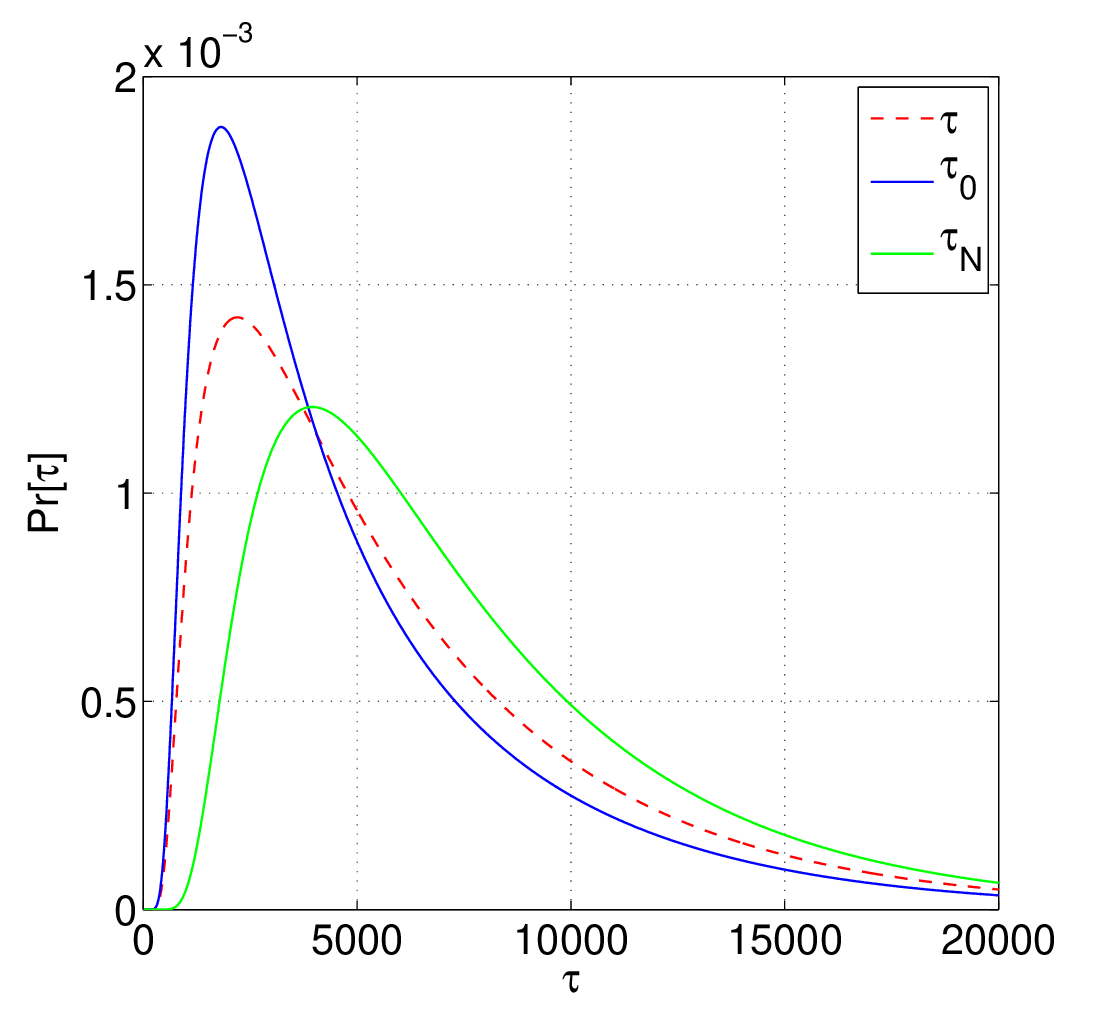}
	\caption{Probability distribution of convergence times for $N =100$ and initial state with $b=33$. Its computation is based on the powers $P^t$ of the transition matrix. Convergence to $b=0$, $\tau_0$, is considerably faster than convergence to $b=N$, $\tau_N$.}
	\label{fig:AbsIndependent}
\end{figure}

Markov chain theory allows for the evaluation of certain properties that are not easily computed with mean--field approaches.
To make one example we computed the mean convergence times to the two consensus states $b=0$ and $b=N$ independently.
This method is described in \cite{Kemeny1976}:64-65 and the basic idea is to "compute all probabilities relative to the hypothesis that the process ends up in the given absorbing state" (ibid:64).
This then leads to a new absorbing chain with the specified state as the only absorbing state.
For convergence to $b = 0$ from an initial configuration with $b$ agents in $\blacksquare$ we obtain
\begin{equation}
\tilde{\tau}^0_b = bN + \sum\limits_{j=b+1}^{N} \frac{b N (N-j)}{j (N-b)}
\label{eq:Converge0}
\end{equation}
and for convergence to $b = N$
\begin{equation}
\tilde{\tau}^N_b = (N-b) N + \sum\limits_{j=1}^{b-1} \frac{j N (N-b)}{b (N-j)}
\label{eq:ConvergeN}
\end{equation}
For a system of 100 agents these times are shown in Fig. \ref{fig:TauIndependent}.
The more the initial situation departs from a fifty--fifty configuration $b = N/2 = 50$, the larger the difference between the two convergence times.
This is also visible in the probability distribution of convergence times shown in Fig. \ref{fig:AbsIndependent} for an initial configuration with $b = 33$.
This shows how the overall convergence behavior is a composite of the two different convergence trends to the two consensus states independently.

\subsection{Further Refinement}

\begin{figure}[hbtp]
\centering
\includegraphics[width=.7\linewidth]{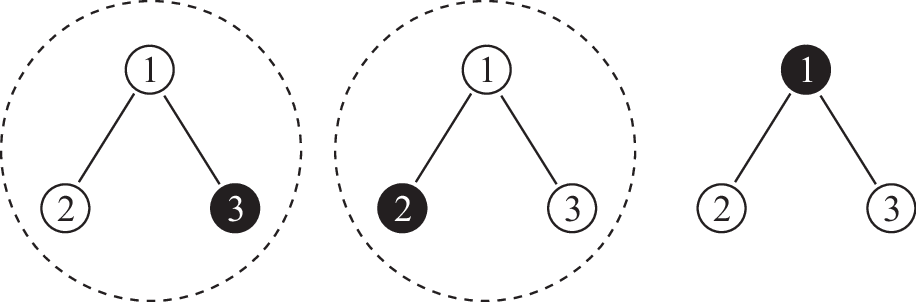}
\caption{The 3 different configurations "g", "f" and "e" of length 3 with one agent in $\blacksquare$ and two in $\square$ ($b = 1$). Only the first two configurations ("f" and "g") are macroscopically equivalent.}
\label{fig:AutomorphismsN3.Configurations}
\end{figure}

Going beyond the complete graph let us first consider again the initial example of three agents shown in Fig.\ref{fig:3Hypercube}.
Let us assume the following interaction probabilities: $\omega(1,2) = \omega(2,1) = \omega(1,3) = \omega(3,1) = \omega$, but $\omega(2,3) = \omega(3,2)=0$ (see Fig. \ref{fig:AutomorphismsN3.Configurations}).
That is, we introduce a slight amount of heterogeneity by cutting the edge between agent 2 and 3.
Let us consider the transitions from $X_{2}$ to $X_3$.
The probability (\ref{eq:PhatVM}) of a transition form configuration "e" to "h" is $\Phat(e,h) = \omega(1,2) + \omega(1,3) = 2\omega$, from "f" to "h" we have $\Phat(f,h) = \omega(2,1) + \omega(2,3) = \omega$ and for  "g" to "h", $\Phat(g,h) = \omega(3,1) + \omega(3,2) = \omega$.
While all these probabilities are equal for the complete graph (as $\omega(i,j) = \omega: \forall i,j$) they are not all equal if one or two connections are absent and so the lumpability condition is violated (in our example $\Phat(e,h) \neq  \Phat(f,h) = \Phat(g,h)$).

Deriving a partition $\mathcal{M}_{\omega}$ such that the micro process projected onto it is a Markov chain requires a refinement of the aggregation procedure.
We use Proposition \ref{propositionlumpability} to identify bundles of micro configurations that can be interchanged without changing the hypercubic micro chain.
The interaction network has a symmetry such that the agents 2 and 3 can be permuted without affecting the connectivity structure, that is, $Aut_{\omega}= (1)(23)$.
This symmetry imposes a symmetry in the hypecubic graph associated to the micro process such that the configurations "f" and "g" with $b = 1$ and respectively "b" and "c" with $b=2$ can be permuted without affecting the transition structure.
In this simple example, therefore, the previous macro atoms $X_2$ (and $X_1$) must be refined such that configurations "g" and "f" ("a" and "b" ) on the one hand and "e" ("c") on the other form different sets in $\mathcal{M}_{\omega}$.

\begin{figure}[hbtp]
\centering
\includegraphics[width=.99\linewidth]{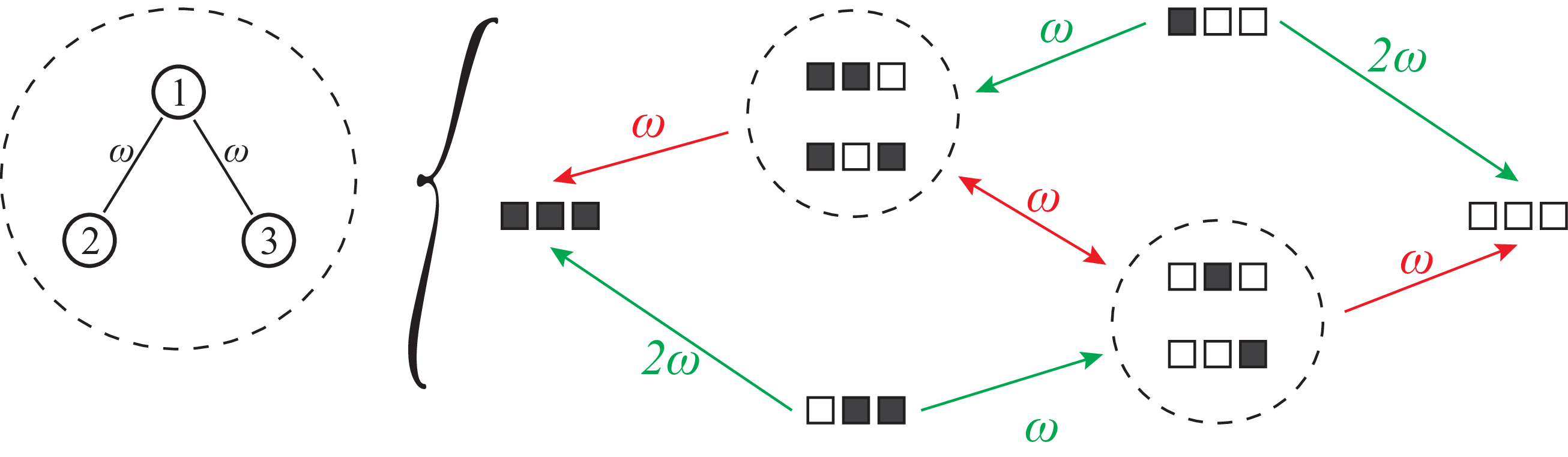}
\caption{Probabilistic structure of the model with three agents if the connection between 2 and 3 is absent.}
\label{fig:VMHypercubeRefinement}
\end{figure}

\subsection{The Two--Community Model}
\label{sec:2Com}

Consider a population composed of two sub-population of size $L$ and $M$ such that $L+M=N$ and assume that individuals within the same sub-population are connected by strong ties whereas only weak ties connect individuals that belong to different communities.
We could think of that in terms of a spatial topology with the paradigmatic example of two villages with intensive interaction among people of the same village and some contact across the villages.
This is similar to the most common interpretation in population genetics where this is called the island model \cite{Wright1943}.
In another reading the model could be related to status homophily \cite{Lazarsfeld1954} accounting for a situation where agents belonging to the same class (social class, race, religious community) interact more intensively than people belonging to different classes.

\begin{figure}[hbtp]
\centering
\includegraphics[width=.7\linewidth]{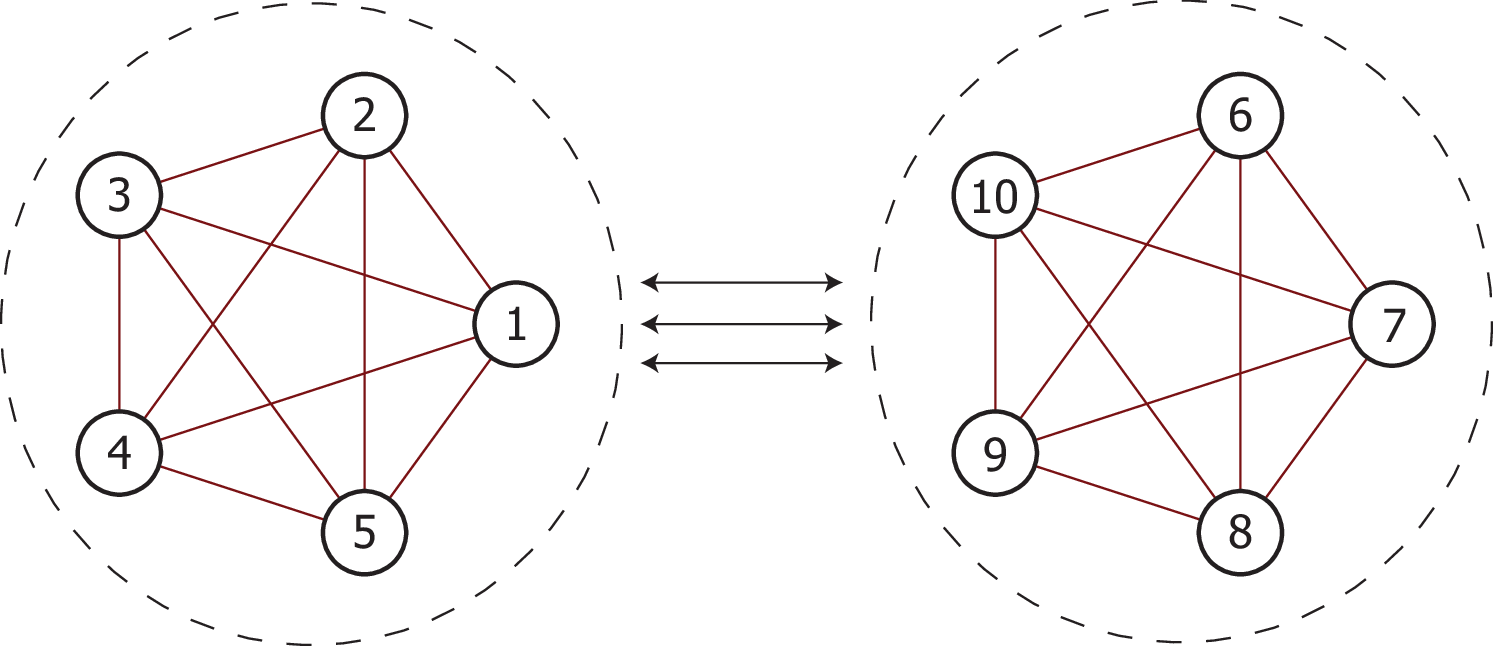}
\caption{A two--component graph with two homogeneous sub--populations.}
\label{fig:TwoPops5}
\end{figure}

Let us adopt the perspective of a weighted graph and say that an edge with weight $w_{ij} = 1$ connects agents of the same community whereas edges across the two communities have a weight $w_{ij} = 1/r$.
Therefore, $r$ is the ratio between strong and weak ties.
For the VM run on such a network, notice that there may be subtle differences in the resulting interaction probabilities $\omega(i,j)$ depending on how the agent choice is performed.
First, in the case of link update dynamics a link $(i,j)$ is chosen out of the set of all links and so the $\omega(i,j)$ are proportional to the edge weight.
Namely, let $\gamma$ denote the interaction probability between agents of the same community and $\alpha$ the respective probability across communities, then
\begin{eqnarray}
\gamma = \frac{r}{2 L M+((L-1) L+(M-1) M) r}\\
\alpha = \frac{1}{2 L M+((L-1) L+(M-1) M) r},
\label{eq:gamma.TwoPops}
\end{eqnarray}
where the divisor is the sum over all edge weights and establishes that  $\sum_{(i,j)} \omega(i,j) = 1$.
A second mode of agent choice is to first choose an agent $i$ and then choose a second agent $j$ out of its neighbor set.
In the case that $M \neq L$, the interaction probabilities become different from (\ref{eq:gamma.TwoPops}).
In the following we will concentrate on the example with $M = L = 50$ so that Eq. (\ref{eq:gamma.TwoPops}) gives the right interaction probabilities for node and link update dynamics.

Notice, moreover, that independent of $M$ and $L$ both update modes give rise to the same symmetry group $Aut_{\omega}(N) =  (1\ldots M)(M+1\ldots N)$.
$Aut_{\omega}(N)$ is composed of the symmetric group $\SG_L$ and $\SG_M$ acting on the two subgraphs and it means that $\omega$ is invariant under permutations of agents within the same community.
Let us denote by $m$ and $l$ the number of $\blacksquare$--agents in $M$ and $L$.
It is then clear that all configurations $x$ and $y$ with $[m(x) = m(y)] \cap [l(x) = l(y)]$ are macroscopically equivalent.
As $0 \leq m \leq M$ and $0 \leq l \leq L$ the aggregation defines a Markov chain with $(M+1)(L+1)$ states which is still very small compared to the number of $2^{(M+L)}$ micro configurations.
Notice that this generalizes naturally to a larger number of subgraphs.
Notice also that the multipartite graphs studied in \cite{Sood2005} fall into this category and that the authors used the respective sub--densities in their mean--field description.

\begin{figure}[h]
	\centering
	\begin{tabular}{c c}
\includegraphics[width=.5\linewidth]{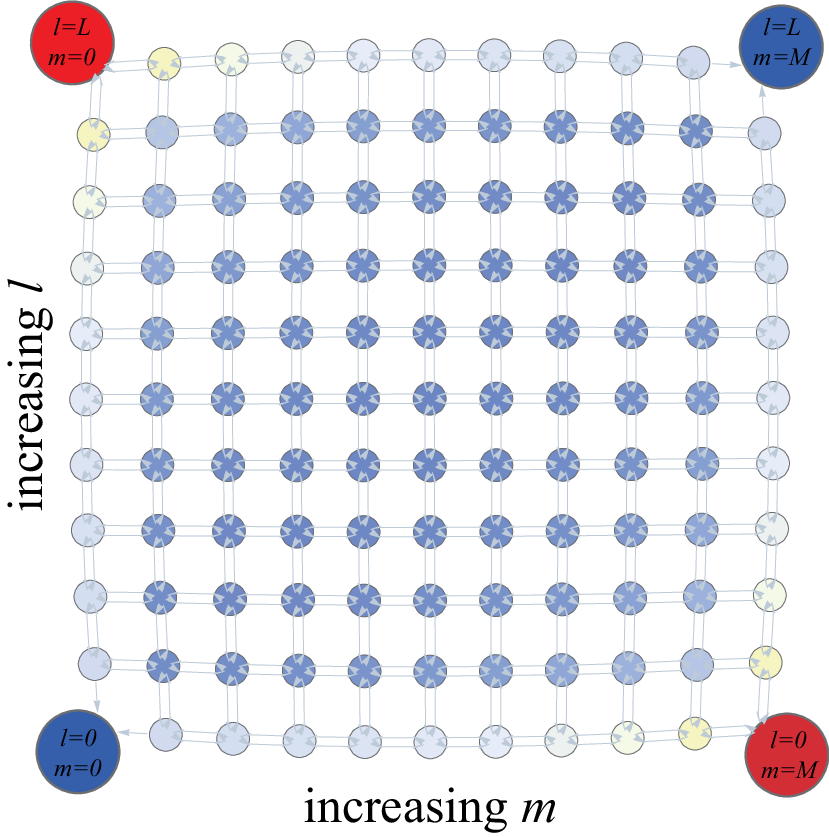} &\includegraphics[width=.5\linewidth]{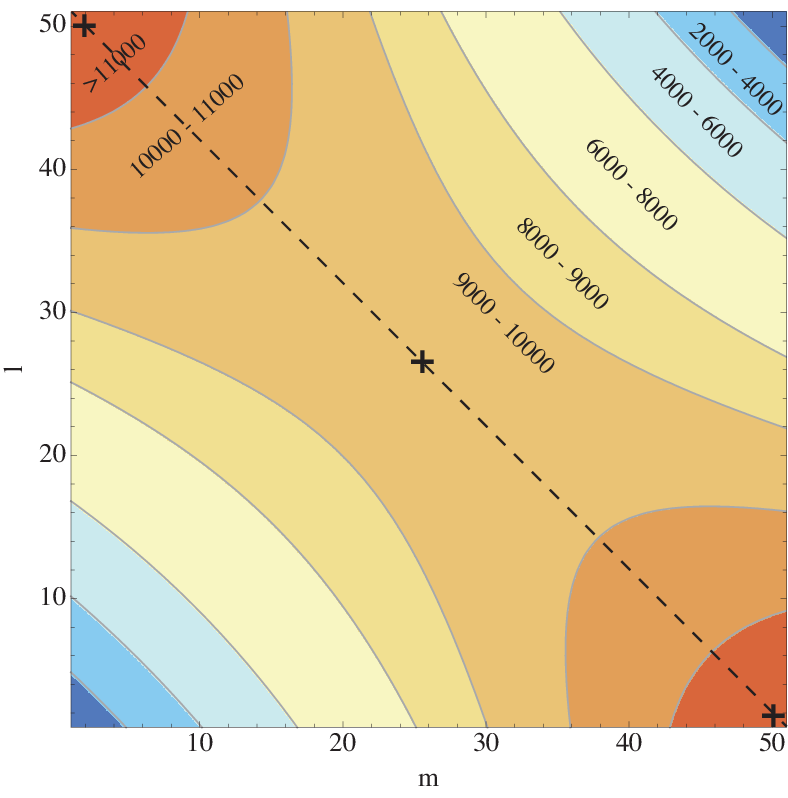}
	\end{tabular}
	\caption{L.h.s.: The structure of the chain for $L=M=10$. The consensus states $\tilde{X}_{0,0}, \tilde{X}_{M,L}$ as well as the states of inter--community polarization $\tilde{X}_{0,L}, \tilde{X}_{M,0}$ are highlighted. The quasi--stationary distribution is mapped into node colors from blue (low values) to red (high values).
	R.h.s.: Mean convergence times $\tau$ for $M = L = 50$ for all initial configurations $\tilde{X}_{m,l}$ for $r = 1/100$. The disordered initial configuration $\tilde{X}_{M/2,L/2}$ and the two partially ordered configurations $\tilde{X}_{M,0}$ and $\tilde{X}_{0,L}$ are highlighted by ${\bf +}$.}
	\label{fig:Tau_Islands_001}
\end{figure}

The structure of the Markov chain associated to the VM on the two--community graph is shown on the l.h.s. of Fig. \ref{fig:Tau_Islands_001}.
For the system of size $M$ and $L$ the transition probabilities for the transitions leaving an atom $\tilde{X}_{m,l}$ are given by
\begin{eqnarray*}
P(\tilde{X}_{m+1,l}|\tilde{X}_{m,l}) & = &\gamma  (m (M-m)) + \alpha  (M-m) l \vspace{6pt}\\
P(\tilde{X}_{m-1,l}|\tilde{X}_{m,l})  &= &\gamma  (m (M-m)) + \alpha  m (L-l)\vspace{6pt}\\
P(\tilde{X}_{m,l+1}|\tilde{X}_{m,l}) &= &\gamma  (L-l) l + \alpha  (L-l)m \\
P(\tilde{X}_{m,l-1}|\tilde{X}_{m,l})  &= &\gamma  (L-l) l + \alpha  (M-m)l
\label{eq:TransProbs.T2.TwoPop}
\end{eqnarray*}
In what follows, we study a system with $M = L = 50$.
This gives a Markov chain of size $(M+1)(L+1) = 2601$.
Notice that the computations (matrix inversion and powers) needed in the analysis of that chain bear already some computational cost and that a further increase in system size will increase these costs greatly.

On the r.h.s. of Fig.\ref{fig:Tau_Islands_001} the mean convergence times are shown for all initial states $\tilde{X}_{m,l}$ and a coupling ratio of $r = 1/100$.
In comparison to the homogeneous mixing case (Eq. \ref{eq:tauk}) the mean number of steps before absorption $\tilde{\tau}_{m,l}$ increases considerably for all initial configurations.
For $m+l = k = 50$ the complete graph will order in average after $6880$ steps whereas this number increases to $9437$ for $m=25, l = 25$.
Notably, it increases further to $11921$ for the initial configurations with consensus within the communities but disagreement across the two islands (inter--community polarization).

\begin{figure}[h]
	\centering
	\begin{tabular}{c c}
\includegraphics[width=.49\linewidth]{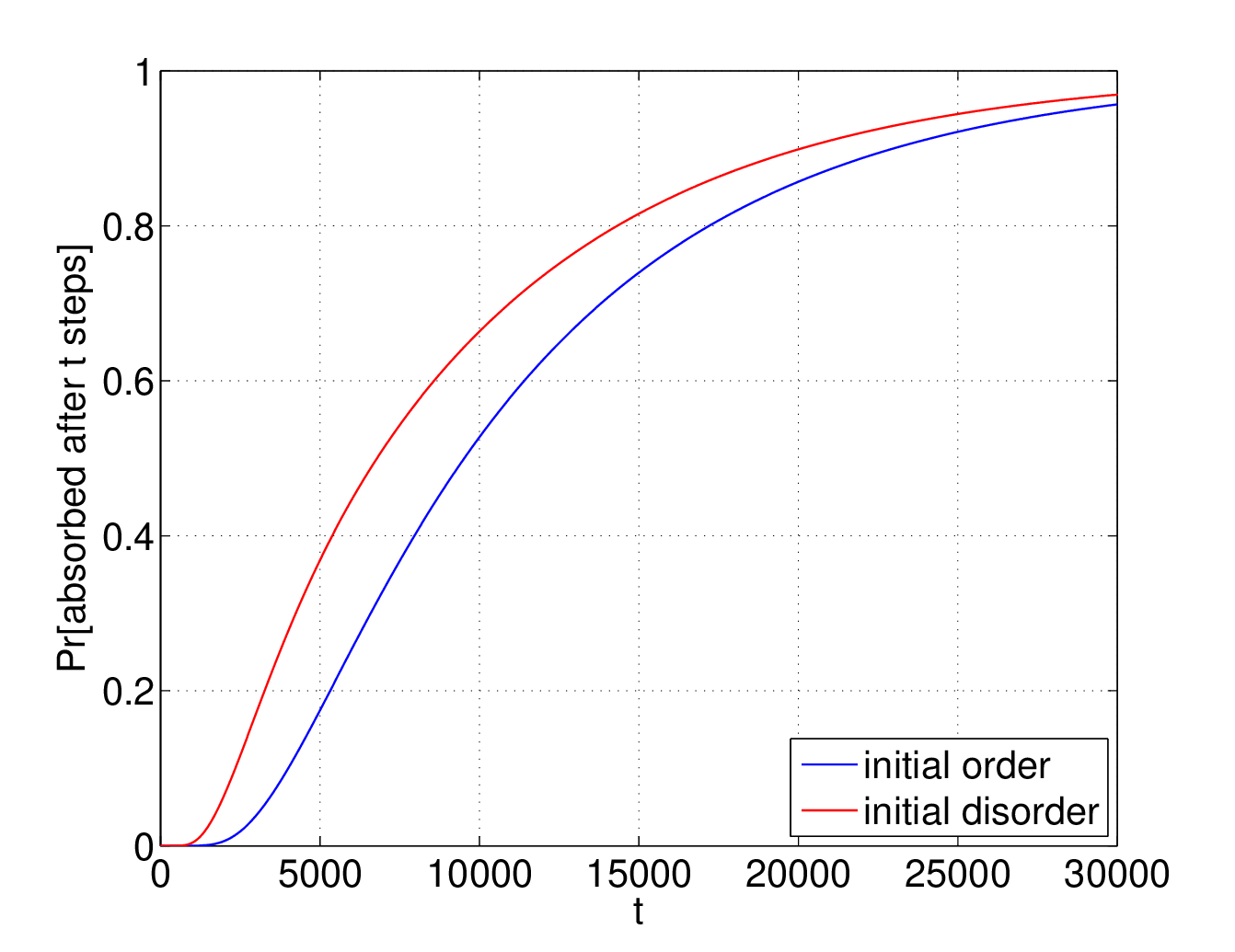}&\includegraphics[width=.49\linewidth]{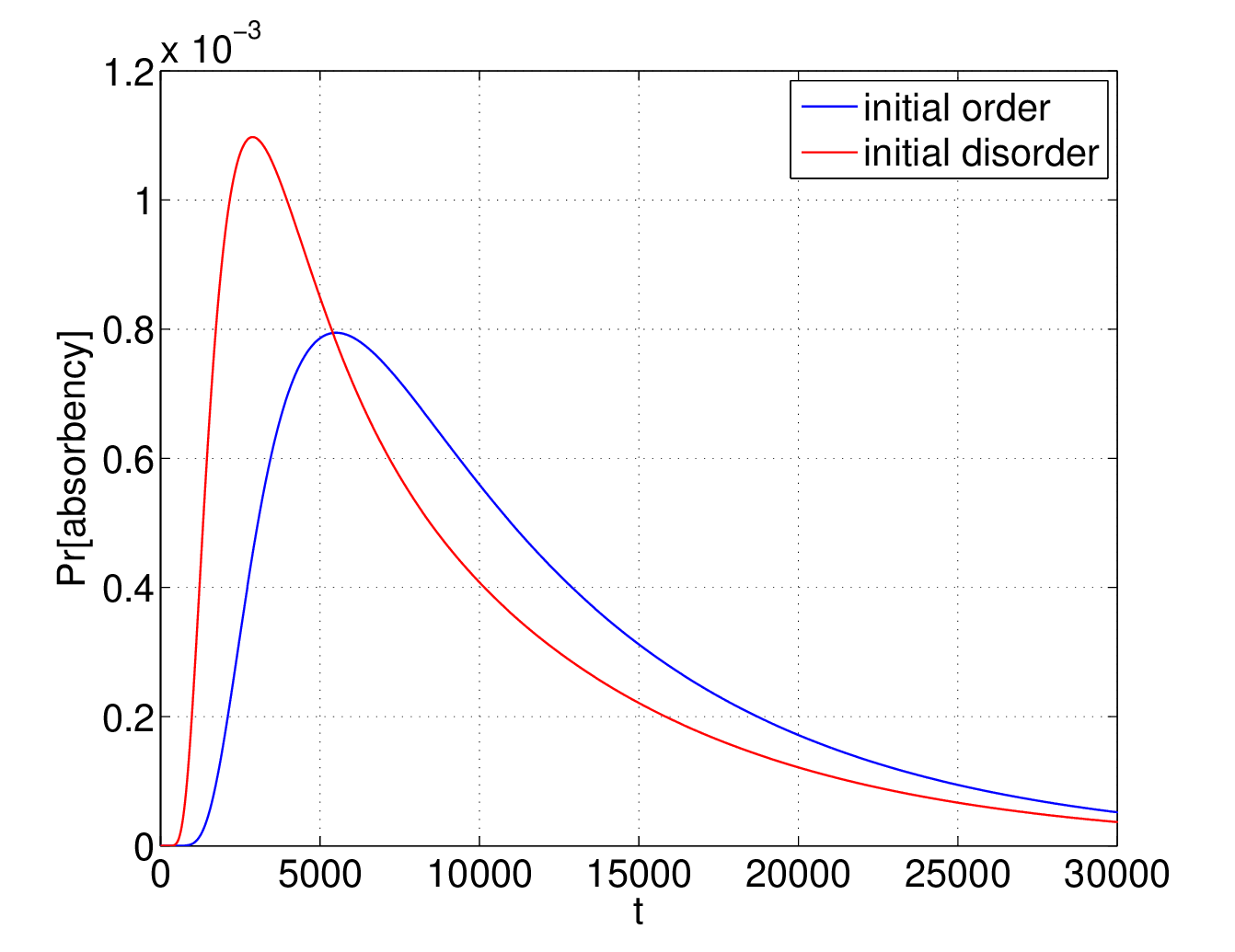}
	\end{tabular}
	\caption{Distribution of convergence times $\tau$ for $M = L = 50$, $\alpha/\gamma = 1/100$ for initial disorder $X_{M/2,L/2}$ (red curves) and initial polarization $\tilde{X}_{M,0}$ and $\tilde{X}_{0,L}$ (blue curves) with consensus among agents of the same island, but disagreement across the two islands. L.h.s.: cumulative probability of being absorbed after $t$ steps- R.h.s.: probability of absorbency at time $t$.}
	\label{fig:cdfAbsorbency.Islands}
\end{figure}

We compare these two situations (namely initial disorder $\tilde{X}_{25,25}$ and initial order in form of inter--community polarization $\tilde{X}_{50,0}$) by considering the distribution of convergence times for two configurations with $m + l = N/2 = 50$.
The respective cumulative distributions for $r= 1/100$ is shown on the l.h.s. of Fig. \ref{fig:cdfAbsorbency.Islands} and on the r.h.s. the respective probability of absorbency at time $t$ is shown.

In the case of initial disorder (red curves), where the states $\square$ and $\blacksquare$ are distributed equally over the two islands, there is a certain number of realizations that approaches one absorbing consensus state without entering the states of partial order ($\tilde{X}_{M,0}$ and $\tilde{X}_{0,L}$).
The probability of absorbency reaches a peak after a relatively short time of around $t \approx 3000$ steps whereas the highest absorbency probability lies around $t \approx 5000$ for the ordered initial condition (blue curves).
At around $t \approx 5000$ already 40 \% of realizations have converged for the disordered case, but only 20 \% in case of initial polarization.
This shows that there is a strong influence of the interaction topology leading to a high heterogeneity in the behavior for different initial configurations with the same global magnetization $k = m+l$.
The ordered configurations $\tilde{X}_{M,0}$ and $\tilde{X}_{0,L}$ function as dynamical traps and it may take a long time to escape from them, especially when the coupling across communities $r$ becomes very weak.
On the other hand, however, Markov chain theory tells us that the probability for very long waiting times decays exponentially.


\begin{figure}[h]
	\centering
\includegraphics[width=.9\linewidth]{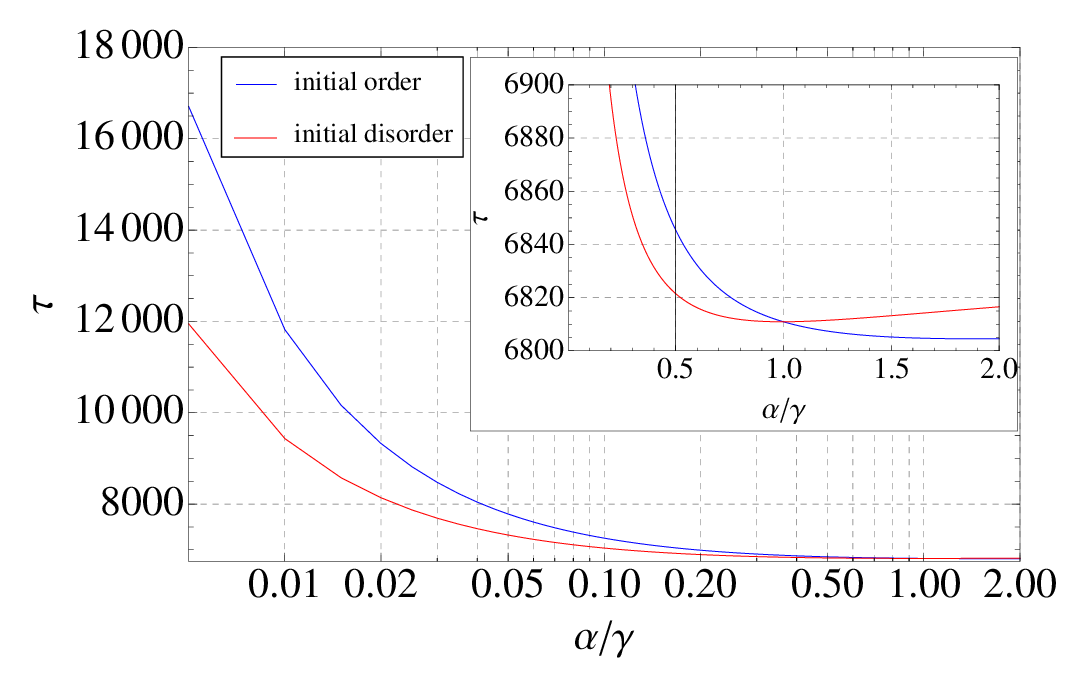}
	\caption{Mean convergence times $\tau$ as a function of the ratio $r = \alpha/\gamma$ between strong and weak ties for the ordered initial configuration $X_{M,0}$ (blue curves) and the disordered initial configuration $X_{M/2,L/2}$ (red curves).}
	\label{fig:tau_VM_Islands_a001_all}
\end{figure}

In Fig. \ref{fig:tau_VM_Islands_a001_all}, a more detailed picture of how convergence times increases as $r = \alpha/\gamma$ decreases is provided.
For the two initial situations considered previously the mean convergence times are shown as a function of the ratio $r = \alpha/\gamma$ between strong and weak ties.
Notice again that these extreme configuration are highlighted by ${\bf +}$ in Fig. \ref{fig:Tau_Islands_001}.
It is clear that the mean times to absorbency diverge as $r$ approaches zero, $\lim\limits_{r \rightarrow 0} \tau = \infty$.
This is due to the fact that the interaction topology becomes disconnected in that extreme case, and therefore, the non-consensus configurations $\tilde{X}_{M,0}$ and $\tilde{X}_{0,L}$ become absorbing.
In other words, to go from (say) $\tilde{X}_{0,L}$ to (say) $\tilde{X}_{0,0}$ requires an infinite number of steps.
In fact, we then deal with a completely new chain that has four absorbing states, or more precisely, with two chains one for each island.
However, as long as $\alpha > 0$ the possibility to escape from $\tilde{X}_{0,L}$ remains, even if it takes very long.

\begin{figure}[htp]
	\centering
	\begin{tabular}{c c}
\includegraphics[width=.5\linewidth]{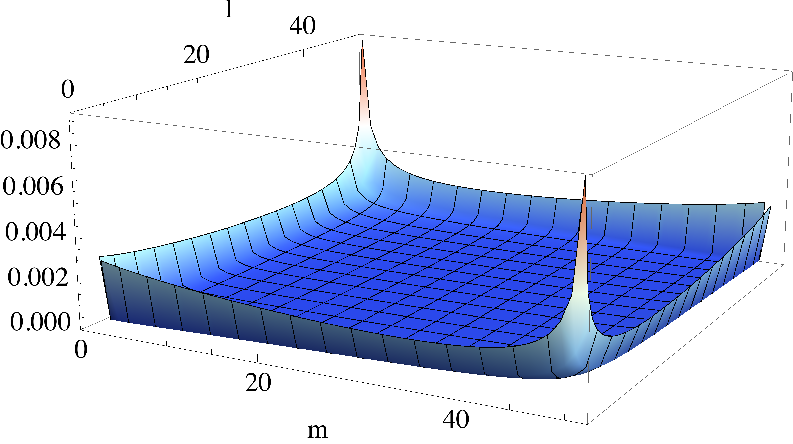}&\includegraphics[width=.5\linewidth]{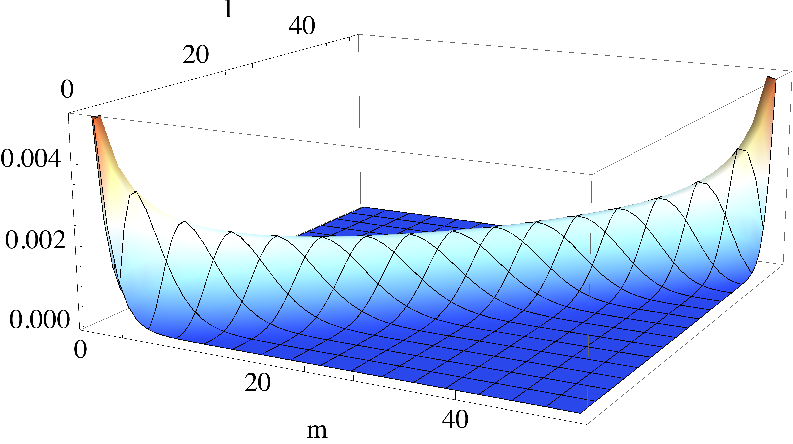}
	\end{tabular}
	\caption{Quasi-stationary distribution for the VM on two islands with $r = 1/100$ (l.h.s.) and $r = 1$ (r.h.s.).}
	\label{fig:QSD_VM_Islands_a001i}
\end{figure}

Finally, to characterize the long--term transient behavior, let us look at the quasi--stationary distribution of the VM.
This distribution contains the probabilities to be in the different transient states for realizations that are not absorbed after a certain time.
It corresponds the normalized left eigenvector associated to the largest eigenvalue of the transient sub--matrix $Q$ of $P$ (just as the stationary distribution of a regular chain is the normalized left eigenvector of the transition matrix $P$).
See, for instance, \cite{Darroch1965} (pages 91 - 93 in particular) for a description of the quasi--stationary distribution.

Fig. \ref{fig:QSD_VM_Islands_a001i} shows the quasi-stationary distribution for the two--community VM with $r = 1/100$ (l.h.s.) and $r = 1$ (r.h.s).
Notice that the latter corresponds to the homogeneous mixing case.
If $r$ is small there is a high (conditional) probability that the process is trapped in one of the states of local order.
Also the states $\tilde{X}_{m,0}$ and $\tilde{X}_{0,l}$ with one uniform sub-population have a relatively high probability indicating that convergence to complete consensus out of local order does not happen via a transition through complete disorder.
This is in stark contrast to the homogeneous mixing situation, which is shown on the r.h.s. of Fig. \ref{fig:QSD_VM_Islands_a001i}.
In this case states of inter--community polarization ($m = M, l = 0$ and $m = 0, l = L$) and states close to that become in effect extremely rare random events.\footnote{The reason for this is clear. The number of micro configurations $x \in \BSigma$ mapped into the state $\tilde{X}_{m,l}$ is $\binom{M}{m}\binom{L}{l}$ which is a huge number for $m \approx M/2, l \approx L/2$ but only 1 for $m = M, l = 0$ and $m = 0, l = L$. Because under homogeneous mixing there is no favoring of particular agent configurations with the same $k = m+l$ the stationary probability at macro scale is proportional to the cardinality of the set $\tilde{X}_{m,l}$.}

\subsection{The Ring}


Prop. \ref{propositionlumpability} generalizes to networks with arbitrary automorphisms which we illustrate at the example of the ring graph.
When the model on the ring with nearest neighbor interactions is defined by $\omega(i,i+1) = \frac{1}{N}: i \bmod N$, it possesses an invariance with respect to translations.
That is, the automorphism group $Aut_{\omega} (N)$ consists of all cyclic shifts of agents generated by $\sigma : (1,2,\ldots,N) \rightarrow  (N,1,2,\ldots,N-1)$. Notice that translational symmetries of this kind also play an important role in the determination of the relevant dimensions of spin rings \cite{Baerwinkel2000} and that there are interesting parallels in between the two problems.

\begin{figure}[hbtp]
\centering
\includegraphics[width=.6\linewidth]{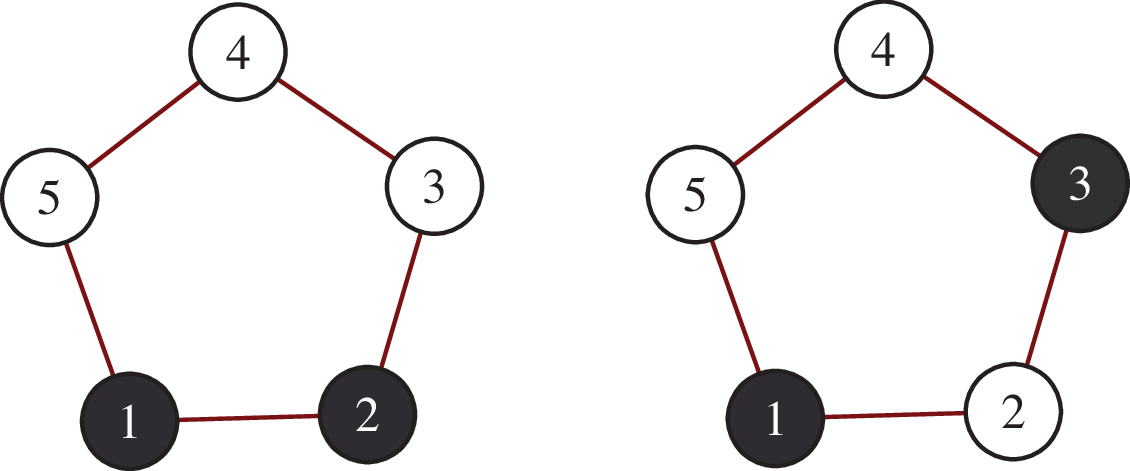}
\caption{Two configurations with an equal number of black agents ($b = 2$) which are not macroscopically equivalent for the ring with $N=5$.}
\label{fig:Ring5.Configurations}
\end{figure}

Consider a ring of five agents ($N = 5$) with $2^5 = 32$ micro states.
For $x = (\blacksquare\blacksquare\blacksquare\blacksquare\blacksquare)$ it is clear that $\sigma^k (x) = x$ for all $k$.
That is,  $x = (\blacksquare\blacksquare\blacksquare\blacksquare\blacksquare)$ with $b=5$ constitutes a class of its own.
For $b=4$, we may use $x_1 = (\square\blacksquare\blacksquare\blacksquare\blacksquare)$ as a generator for its class, Eq.(\ref{eq:construction}).
As all 5 configurations with $b=4$ can be obtained shifting $x_1$, all of them are in the same equivalence class.
The 10 configurations with $b = 3$ cannot be lumped into the same macro state.
There are two classes differentiated by the distance of zero or one in between the two black agents (see Fig. \ref{fig:Ring5.Configurations}).
Using the two configurations shown in Fig. \ref{fig:Ring5.Configurations} as generators  yields two equivalence classes each containing five micro states.
The cases $b = 2,1,0$ follow by symmetry so that all in all the dimension $\X$ of macro chain is reduced to 8.

In the general case of $N$ agents we can in principle proceed in the same way.
However, the number of macro states will increase considerably with the system size.
We finish this section with a quantification of this number for the ring for which we can use a well--known enumeration theorem due to P\'{o}lya (see \cite{Harary1973}:35-45, Eqs.(2.2.10) and (2.4.15) in particular).
According to this, the number of macro states is 
\begin{equation}
	|\X| = \frac{1}{N} \sum\limits_{k|N}^{} \varphi(k) 2^{\frac{N}{k}}
	\label{eq:polya}
\end{equation}
where $\varphi(k)$ is the Euler $\varphi$--function and the sum is over the divisors $k|N$ of $N$.
As an approximation (in fact, a lower bound) we have $|\X| \approx 2^N/N$.
Hence, an explicit solution of the macro chain will be possible only for very small systems.

\section{Scope and Limitations}

The method described in this paper applies not only to the VM.
In fact, the micro formulation and Prop. \ref{propositionlumpability} can be applied without modification to any interacting particle system in which the local transition probabilities are a function solely of the local neighborhood configuration, as defined by an unchanging graph.\footnote{We thank one anonymous reviewer for this comment.}
As shown in \cite{Banisch2012son}, the method is not restricted to binary agent attributes $x_i \in \{\square,\blacksquare\}$, but can in principle be applied to any agent model with a finite set of agent attributes.
It has been used in \cite{Banisch2012dnc} in the context of evolutionary processes to compare different update schemes and analyze their effects on adaptation and speciation.
Possible future application examples include stochastic cellular automata on graphs with asynchronous update, other models of opinion and socio--cultural dynamics (e.g., \cite{Axelrod1997,Banisch2010acs}), as well as spin--based models of market dynamics (\cite{Bornholdt2001,Krause2011}).

\rem{
The method is clearly not restricted to binary agent attributes $x_i \in \{\square,\blacksquare\}$, but can in principle be applied to any agent model with a finite set of agent attributes.
Whenever only one agent (say $i$) is allowed to change its state at a time and the update depends only on the current configuration, the agent model performs a random walk on the graph defined by the adjacency relation $x \stackrel{i}{\sim}  y$ (notice that $P(x,y) > 0 \Rightarrow x \stackrel{i}{\sim}  y$ but not vice versa).
If $\delta$ is the number of agent attributes, then the graph associated to the micro process is the Hamming graph $H(N,\delta)$ (with loops), which is known to possess a relatively large symmetry group.
This means that the number of $\hat{\sigma}$ with $P(x,y) = P(\hat{\sigma}(x),\hat{\sigma}(y))$ may be large 
Under the hypothesis of local update on an unchanging graph Prop. \ref{propositionlumpability} can be used to find such a group of $\hat{\sigma}$.
}

However, the more complex the internal structure of the agents and the more heterogeneous their interaction behavior, the lower our chances to derive a loss--less coarse--graining that leads to a tractable Markov chain.
It is clear that in heterogeneous networks with a small number of automorphisms the coarse--graining is limited because only a few micro states are macroscopically equivalent and can be lumped.
The ring graph (with invariance to cyclic shifts) gives an idea about the limitations of the approach.
As there are $N$ non-trivial automorphisms for a ring of size $N$, the number of macro states in a Markovian coarse-graining is bound by $2^N/N$.
This certainly reduces the problem and may be useful for small systems, but its practical use for large $N$ is limited.
As this method is based on exact graph automorphisms it is more suited for stylized situations as the two--community model discussed in Section \ref{examples}.
We envision the application of Prop. \ref{propositionlumpability} to some interesting hierarchical compositions of sub-populations such as a relatively small number of communities placed on a ring or a lattice.

On the other hand, the method informs us in this way about the complexity of a system introduced by non--trivial interaction relations.
Even in a model as simple as the VM, the behavior of whole system is not completely described by summation over its elements (aggregation in terms of $b$), because non--trivial dynamical and spatial effects may emerge at the macro level.
In this sense, our work is related to key concepts in the area of computational emergence (\cite{Bedau2002,Huneman2008}) dealing with criteria and proper definitions of emergence.
Thereafter "an emergent phenomenon is one that arises from a computationally incompressible process" (\cite{Huneman2008}: 425/26).
Markov projections as discussed here in the context of the VM provide explicit knowledge about the (in)compressibility of computational models and may therefore help to operationalize these rather abstract definitions.

For its relation to emergence and complexity, the case in which Markovianity is lost in the transition from micro to macro is, in some sense, even more interesting than the lumpable case.
Namely because it is more relevant in the setting of general ABMs where Markovianity is usually lost in the transition to the desired level of observation.
In order to apply Markov chain aggregation in such a more general setting, a first crucial step consists of a rigorous characterization of the macro-level effects that are introduced by aggregation without sensitivity to microscopic details.
This could be based on information-theoretic measures that quantify deviations from Markovianity (see \cite{Goernerup2008} and \cite{Banisch2014phd}, Chapter 5, in particular).
Such a characterization of "macroscopic complexity" -- understood as a quantification of deviations from Markovianity -- might provide a suitable framework to evaluate approximate macro descriptions and enable a systematic analysis of order parameters and their adequateness to describe the respective model dynamics.
It may also help to identify the role of certain micro-structural patterns in the creation of complex macroscopic outcomes such as non-trivial temporal correlations.

Let us finally note that in general there may be many partitions $\mathcal{M}$ of the state space that are lumpable and here no statement is made here about optimality of the partition $\mathcal{M}_{\omega}$ generated by the application of Prop. \ref{propositionlumpability}.
On the other hand, a simple answer  is provided by a closer inspection of the VM with homogeneous mixing telling us that $\mathcal{M}_{\omega} = \X$ is not optimal in that case.
Namely, we have for any $b$, $P(X_{b \pm 1}|X_b) = P(X_{(N-b) \mp 1}|X_{(N-b)})$ which means that the pairs $\{X_b,X_{(N-b)}\}$ can be lumped into the same state.
In other words, the macro chain on $\X$ is lumpable with respect to an even coarser description.
The reason for this is that the VM update rule brings about an additional symmetry that is not accounted for in $Aut_{\omega}$ and therefore not in $\mathcal{M}_{\omega}$.
More generally, the micro structure of the VM is always symmetric with respect to the simultaneous flip of all agent states $x_i \rightarrow \bar{x}_i, \forall i$ and therefore, independent of the interaction topology, $\Phat(x,y) = \Phat(\bar{x},\bar{y})$. 

\section{Conclusion}

In conclusion, this paper describes a way to reduce the state space of the VM by exploiting all the dynamical redundancies that have its source in the agent network.
In the aggregation of micro states into macro atoms following this proposal no information about the details of the dynamical process is omitted.
However, in heterogeneous interaction substrates with a small number of automorphisms exact coarse--graining is limited because only a few micro states are macroscopically equivalent and can be lumped.
On the other hand, whenever it is possible to derive a tractable Markov chain at the macro level, Markov chain theory provides us with tools that allow for a complete understanding of the model dynamics.

The probabilistic setting we adopt allows to relate microscopic agent dynamics to the macro evolution of aggregate observable variables and 
shows how network heterogeneities translate into heterogeneities in the dynamical structure of the model.
A characterization of the dynamical effects that may emerge at the macro level will be addressed by future work.

\vspace{6pt}
\small
\subsection*{Acknowledgments}
The authors are grateful to Philippe Blanchard, Edgardo Ugalde and Dima Volchenkov for helpful discussions. SB acknowledges financial support of the German Federal Ministry of Education and Research (BMBF) through the project \emph{Linguistic Networks} ({\tt http://project.linguistic-networks.net}) and the European Community's Seventh Framework Programme (FP7/2007-2013) under grant agreement no.~318723 (MatheMACS).



\end{document}